\documentclass[12pt]{spieman}
\usepackage{amsmath,amsfonts,amssymb}
\usepackage{graphicx}
\usepackage{setspace}
\usepackage{tocloft}
\usepackage{breqn}
\usepackage{algorithmic}
\usepackage{array}
\usepackage{url}
\usepackage{verbatim}
\usepackage{color}
\usepackage{booktabs}
\usepackage[para]{threeparttable}
\usepackage{xcolor}
\usepackage{ulem}
\usepackage{subcaption}
\usepackage{listings}
\usepackage{matlab-prettifier}
\usepackage[pagewise]{lineno}

\lstnewenvironment{python}[1][]{
    \lstset{
        language=Python,
        basicstyle=\ttfamily\small,
        frame=single,
        aboveskip=0.5 pt,
        belowskip=0.5 pt,
        #1
    }
}{}

\title{A sixteen multiple-amplifier-sensing CCD and characterization techniques targeting the next generation of astronomical instruments}

\author[a, b, c,*]{Agust\'in J. Lapi}
\author[a, c, d]{Blas J. Irigoyen Gimenez}
\author[a]{Miqueas E. Gamero}
\author[a, c]{Claudio R. Chavez Blanco}
\author[a, b]{Fernando Chierchie}
\author[c, e]{Guillermo Fernandez Moroni}
\author[f]{Stephen Holland}
\author[c, g]{Ana M. Botti}
\author[c]{Brenda A. Cervantes-Vergara}
\author[c]{Javier Tiffenberg}
\author[c]{Juan Estrada}

\affil[a]{Departamento de Ingenier\'ia El\'ectrica y de Computadoras (DIEC), Universidad Nacional del Sur (UNS), Bah\'ia Blanca, 8000, Argentina}
\affil[b]{Instituto de Inv. en Ing. El\'ectrica ``Alfredo Desages'' (IIIE), CONICET, Bah\'ia Blanca, 8000, Argentina}
\affil[c]{Fermi National Accelerator Laboratory, P.O. Box 500, Batavia, Il 60510, USA}
\affil[d]{Facultad de Ingenier\'ia, Universidad Nacional de Asunci\'on (UNA), San Lorenzo, Paraguay}
\affil[e]{Department of Astronomy and Astrophysics, University of Chicago, 5640 South Ellis Avenue, Chicago, IL, 60637, USA}
\affil[f]{Lawrence Berkeley National Laboratory, One Cyclotron Rd, Berkeley, CA 94720, USA}
\affil[g]{Kavli Institute for Cosmological Physics, University of Chicago, Chicago, IL, 60637, USA}

\cftpagenumbersoff{figure}
\cftpagenumbersoff{table} 

\DeclareMathOperator{\var}{Var}

\begin{document} 
\maketitle

\begin{abstract}
This work presents a candidate sensor for future spectroscopic applications, such as a Stage-5 Spectroscopic Survey Experiment or the Habitable Worlds Observatory. This new type of CCD sensor features multiple in-line amplifiers at its output stage allowing multiple measurements of the same charge packet, either in each amplifier and/or in the different amplifiers. Recently, the operation of an 8-amplifier sensor has been experimentally demonstrated, and the operation of a 16-amplifier sensor is presented in this work. This new sensor enables a noise level of approximately 1\,$e_{rms}^-$ with a single sample per amplifier. Additionally, it is shown that sub-electron noise can be achieved using multiple samples per amplifier.
In addition to demonstrating the performance of the 16-amplifier sensor, this work aims to create a framework for future analysis and performance optimization of this type of detectors. New models and techniques are presented to characterize specific parameters, which are absent in conventional CCDs and Skipper-CCDs: charge transfer between amplifiers and independent and common noise in the amplifiers, and their processing.
\end{abstract}


{\noindent \footnotesize\textbf{*}Agust\'in J. Lapi,  \linkable{lapiagustinjavier@gmail.com} }

\begin{spacing}{2}  

\section{Introduction}

Low-noise silicon imagers have been identified as a key technology for the construction of the next generation of scientific spectroscopic experiments. The low light signal projected on the pixels due to the spectral dispersion on faint objects could be obscured by the uncertainty added by different background sources. In particular, for terrestrial observations, the readout noise of the sensor could be a considerable contribution in the blue region of the spectrum where the sky background is suppressed. For example, this regime is particularly relevant for the planned Stage-5 Spectroscopic Survey Experiment (Spec-S5) \cite{Schlegel:2022}, which would measure $\sim10^8$ distant galaxies (an order of magnitude more than current surveys) for redshift from 2 to 5 times the original wavelength to study the mechanism driving the expansion of the universe after inflation.

Recent studies in \cite{Drlica_2020} show that the signal-to-noise ratio of this spectral line can be increased using the non-destructive readout of Skipper Charge Coupled Devices (Skipper-CCD) to reduce the readout noise contribution in the pixel measurement. At the same time, preliminary studies presented as part of the particle physics community future planning in the US (known as the Snowmass process) in \cite{p5_desi_2, p5_stage_5} show that the rate of the successful measurement of the redshift of Lyman-Break galaxies can be increased by approximately 25\% by reducing the readout noise down to 1e$^-$ compared to the RMS noise level of around 3e$^-$ in the Dark Energy Survey Instrument \cite{desicollaboration2016desi}. This opens an opportunity to increase the survey speed of future terrestrial spectroscopic surveys by improving the readout noise of the detectors.

Another spectroscopic application seeking low-noise silicon sensor technology is the search for Earth-like planets in the habitable zones of Sun-like stars using a coronagraph instrument in space \cite{RauscherNASA2022}. In this case, the visible and near-infrared bands contain abundant information about exoplanet atmospheres. The low expected flux, in the order of a few photons per hour per pixel \cite{Rauscher_photon_rate}, requires sub-electron noise for single-photon detection and fast readout (total exposure time of around 60 seconds) to avoid excessive occupancy of cosmic ray traces in the sensor. For this case, the non-destructive readout of the Skipper-CCD\cite{Tiffenberg:2017aac}, has been identified as a candidate solution.

The Skipper-CCD provides a powerful way, by multiple measurements of the collected charge, to reduce the readout noise of the pixel at the expense of an extra read time due to the multiple sampling which is not tolerated for this kind of application. An extended version of it, called the Multiple-Amplifier-Sensing Charge Coupled Device (MAS-CCD), was recently presented in \cite{holland_2023} and its principles demonstrated in \cite{botti2023fast} provides a solution to overcome the extra read time of Skipper-CCDs. Its good performance encouraged further characterization efforts presented in \cite{MAS_blas_SPIE, ken_spie, ken_24}. The multiple inline architecture of the MAS-CCD, as shown in the simplified schematic of Fig.~\ref{fig:output_stage}\textcolor{blue}{a}, measures the pixel charge packet sequentially, and the final pixel value is computed using the multiple non-destructive measurements taken. This provides an interesting solution to meet the requirement without increasing the total read time compared to the current devices used in the current experiments \cite{desicollaboration2016desi}.
This paper extends the results in \ref{exp_results} to a detector of sixteen output stages which allows reaching the noise operation regime expected for Spec-S5. 
At the same time, the paper provides a new theoretical framework, techniques, and tools to characterize and optimize the new features of the sensor compared to regular CCDs. In particular, the article provides the method to optimally mix the information from the multiple amplifiers and gives a model for a new source of charge transfer inefficiency, which we call node removal inefficiency, related to the extraction from the sense node back to the serial register pixels.

In the following section, an introduction to MAS architecture is addressed. In Sec. \ref{data_procesin} digital signal processing and optimum averaging techniques are presented to further improve the detector performance. A mathematical framework to model Node Removal Efficiency in MAS detectors is addressed in Sec. \ref{sec:gre}. Experimental results are shown in Sec. \ref{exp_results}. Finally, concluding remarks are given in Sec. \ref{sec:conclusion}.

\section{MAS-CCD readout technique and noise performance analysis}
\begin{figure}[t!]
    \centering
    \includegraphics[width=0.8\linewidth]{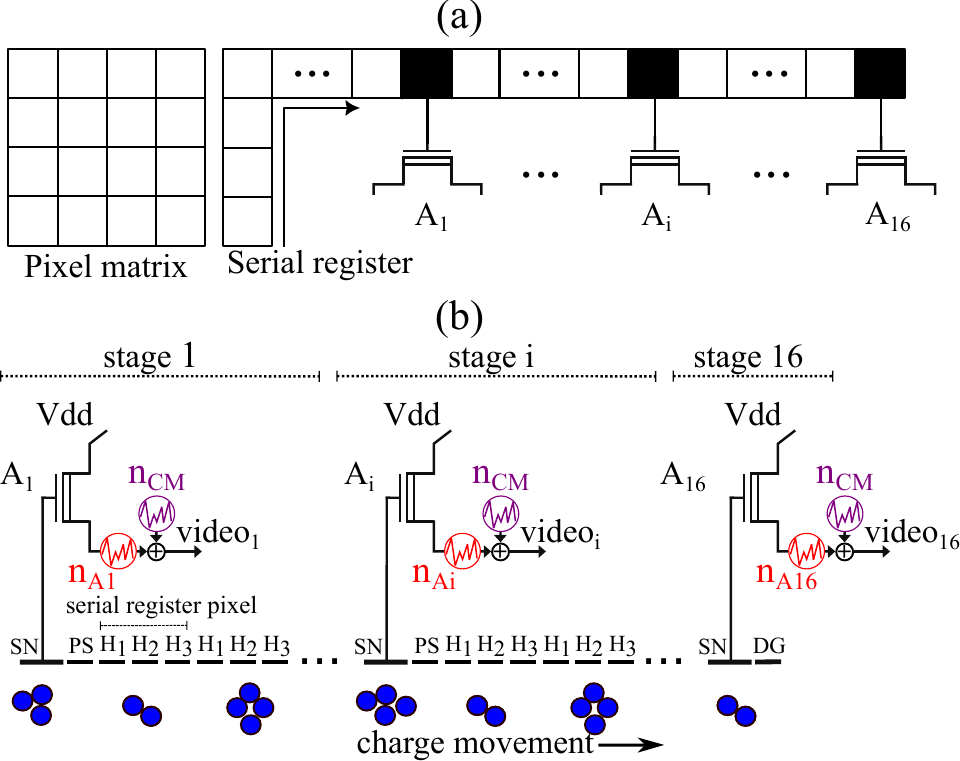}
    \caption{a) Simplified MAS-CCD schematic, conformed by the pixel matrix, a 90-degree bent serial register, and A$_i$ inline amplifiers. b) A more detailed schematic of the MAS-CCD output stage showing its operation and a model of the noise source.}
    \label{fig:output_stage}
\end{figure}

Figure \ref{fig:output_stage}\textcolor{blue}{a} shows a MAS-CCD simplified schematic, its basic components are the pixel matrix, a bent serial register, and 16 inline amplifiers. In a basic readout scheme, the pixel charge packets are transferred from the matrix to the serial register, and then transferred and readout across the multi amplifiers stages (A$_1$ to A$_{16}$).

Figure \ref{fig:output_stage}\textcolor{blue}{b} presents a simplified view of the components and gates of the readout stages at the end of the serial register of the sensor. For each stage, there is a sense node (SN) connected to the gate of the transistors (A$_i$) to measure the charge in the serial register channel. The pixel separation gate (PS) is used to remove the charge from the SN after its readout to put it under the horizontal gates (H1, H2, and H3). These three gates form the structure for one pixel in the serial register, where the rest of the charge is stored. A dump gate is located at the end of the last amplifier for charge disposal, for a more detailed description refer to \cite{holland_2023, botti2023fast}.

In this architecture, the individual amplifiers are capacitively connected to the channel of the sensor for non-destructive readout operation. Each amplifier can take several samples of the pixel charge and move the charge forward to allow its measurement by the next transistor. In the simplest calculation, the final pixel value is computed as the average of the available measurements of the pixel charge
\begin{equation}
\label{eq:average equal weights}
    \text{pixel value} = \frac{1}{N_{a}}\frac{1}{N_s}\sum^{N_{a}}_{i=1}\sum^{N_{s}}_{j=1}s_{i,j},
\end{equation}
 where $s_{i,j}$ is the $j$ sample from the amplifier $i$, and $N_s$ and $N_a$ are the total number of samples taken by each amplifier and the total number of amplifiers, respectively. The readout uncertainty contribution is reduced as the square root of the number of samples $N_s$ as demonstrated for the Skipper-CCD \cite{Tiffenberg:2017aac, skipper_2012, cancelo2021low}. With the MAS-CCD, the noise of each amplifier can be modeled as $n_i=n_{Ai}+n_{CM}$ with $n_{Ai}$ the independent noise of each amplifier and $n_{CM}$ the common noise source affecting all the amplifier stages, as shown in Fig.~\ref{fig:output_stage}\textcolor{blue}{b}. As explained in the following section, there are tools that can be used to monitor and reduce the effect of this correlated noise in the final readout noise performance of the sensor. For the calculations presented in this article the pixel values $s_{i,j}$ are computed using the dual-slope integrator (DSI) processing technique \cite[Chapter~6, pp.537-541]{janesick2001scientific}.
 
When $n_{CM}\ll n_{Ai}$ (or $n_{CM}=0$) the MAS readout noise is also reduced by the square root of $N_a$ \cite{botti2023fast}. When the noise from the different amplifiers is independent but not necessarily equal, the noise uncertainty in the final pixel value is

 \begin{equation}
\label{eq:expected noise}
    \sigma_{ind} =\dfrac{\sqrt{\sum_{i=1}^{N_a}\sigma_{i}^2}}{N_a}
\end{equation} 
where $\sigma_{ind}$ is the readout noise standard deviation of the final value of the pixel and $\sigma_{i}$ is the standard deviation of the noise of the channel $i$ after averaging all its available samples. 

 When the common noise contribution is not zero, as in the case of noise introduced by common power supplies and clock voltage, the final pixel value uncertainty is a function that depends on the effect of the noise in all the channels at the same time. The total uncertainty 
 
\begin{equation}
\sigma_{pix} = \sqrt{\sigma_{ind}^2 + \sigma^2_{n_{CM}}}    
\end{equation}
where $\sigma^2_{n_{CM}}$ is the variance contribution common mode noise at the output of the MAS. $\sigma^2_{n_{CM}}$ can be evaluated using the power spectrum of the common noise in the system and the frequency response of the MAS-CCD readout operation. The frequency response of the MAS-CCD is calculated in Appendix \ref{appendix_freq_response} and compared to the frequency responses of the single and multiple DSI of the Skipper-CCD.

\section{Data processing for optimal noise performance}

\label{data_procesin}
Direct averaging the samples available from the different amplifiers may not be the optimal strategy to get the final pixel values with the best possible noise in all scenarios. Factors such as different noise performance in the amplifiers, different gains, and the presence of common noise in the video signals encourage the use of different weights to combine this information. In the following subsections, the techniques and procedures to implement this idea are developed. The first one develops a technique to equalize the gain of the images coming out from each amplifier. The second subsection provides the methodology to optimally mix the channels based on the equalized images. The third subsection provides a more aggressive approach, where the correlated noise contribution in a channel can be reduced using available information from the other channels.

\subsection{Gain equalization}
\label{sec:gain_equalization}
The impact of the readout noise on signal measurements is related to the noise of the channels, and therefore, it should be incorporated as one of the optimizing aspects for combining the output data from different amplifiers. 

Typically, combining information from different amplifiers requires absolute calibration of each output stage. The MAS-CCD provides a way to circumvent this process by comparing the measurement of the same charge packet by all the amplifiers. As will be seen in the following sections, the optimum combination of the information from different channels only requires the relative gain calibration across the channel. We define the equalization coefficient for the $i$-th channel ($c_i$), which measures the ratio of the gain of the $i$ amplifier relative to the average gain of all the amplifiers, as

\begin{equation}
    c_i = \frac{1}{s_i N_a} \sum_{k=1}^{N_a} s_k
    \label{eq:equalization}
\end{equation}
where $s_k$ is the sample value of a charged pixel from the $k$-th amplifier. $c_i$ measures the ratio of the gain of the $i$ amplifier relative to the average gain of all the amplifiers. The pixel value $s_k$ must be much larger than the total uncertainty ($s_k>>\sigma_{pix}$), but small enough to avoid saturated values.

If many samples are taken on each amplifier, it is assumed that they are averaged before this calculation: $s_k=(1/N_s)\sum_{j=1}^{N_s}s_{k,j}$. 
For the next subsections, it is assumed that the images are equalized as $\tilde{s}_i = s_i/c_i$.

\subsection{Optimum average of the channels}
\label{Sec:optimizedAverge}

Assuming that each amplifier can be combined using different weights, the final pixel value using the equalized measurements $\tilde{s}_i$ of each of the $N_a$ amplifiers can be calculated as

\begin{equation}
    \hat{s} = \sum_{i=1}^{N_a} \alpha_{i} \tilde{s}_{i}
    \label{eq:optAverage}
\end{equation}
where $\alpha_i$ is the corresponding weight of the $i$-th amplifier. These values are calculated so that the variance of the pixel values due to readout noise is minimized. Then, the weights $\alpha_{i}$ are obtained by minimizing 
\begin{equation}
    \var(\hat{s}) = A^T \Sigma A
\end{equation}
where $A^T = [\alpha_{1}, ..., \alpha_{N_a}]$ and 
\begin{equation}
\label{eq:covariance matrix}
    \Sigma = 
    \begin{bmatrix}
    \sigma_1^2 & \cdots & \sigma_{1,N_a}^2 \\
    \vdots & \ddots & \vdots \\
    \sigma_{N_a,1}^2 & \cdots & \sigma_{N_a}^2
    \end{bmatrix}
\end{equation}
is the covariance matrix ($\sigma_{n,m}^2 = \text{cov}(\tilde{s}_n,\tilde{s}_m)$). The common noise from the amplifiers will be reflected as non-diagonal values different from zero in the matrix. Assuming that the noises are stationary, these values can be obtained from the output images using overscan pixels. 

A restriction must be set using Lagrangian multipliers to avoid the trivial solution. Thus, the minimization problem can be stated as
\begin{equation*}
    \text{min}_A \ A^T \Sigma A, \text{subject to} \ A^T \mathbf{1}= 1
\end{equation*}
where $\mathbf{1}$ represents a column vector of ones and dimensions $N_a \times 1$ and the condition is set so that the sum of the $\alpha_i$ coefficients is 1. Weights are then obtained solving the problem for the vector of weights $A$, resulting in
\begin{equation*}
    A = (\mathbf{1}^T\Sigma^{-1}\mathbf{1})^{-1} \Sigma^{-1} \mathbf{1}.
\end{equation*}
For example, solving this equation to find a generic solution for $A^T=[\alpha_1,\alpha_2]$ (assuming a $\Sigma$ matrix of order 2 ($N_a =2$)) leads to $\alpha_1 = (\sigma_{2,2}^2-\sigma_{c}^2)/(\sigma_{2,2}^2+\sigma_{1,1}^2-2\sigma_{c}^2)$ and $\alpha_2 = (\sigma_{1,1}^2-\sigma_{c}^2)/(\sigma_{2,2}^2+\sigma_{1,1}^2-2\sigma_{c}^2)$, where $\sigma_{c}=\sigma_{1,2}=\sigma_{2,1}$. 
From here, it is possible to evaluate different conditions.
If $\sigma_{c}^2 \ll \sigma_{2,2}^2$ and $\sigma_{c}^2\ll \sigma_{1,1}^2$, therefore $\alpha_1 \approx \sigma_{2,2}^2/(\sigma_{2,2}^2+\sigma_{1,1}^2)$ and $\alpha_2 \approx \sigma_{1,1}^2/(\sigma_{2,2}^2+\sigma_{1,1}^2)$.

Considering equal contributions of noise $\sigma_{2,2}^2=\sigma_{1,1}^2$ leads to $\alpha_1=\alpha_2=1/2$. Another option is to consider different contribution of noise $\sigma_{2,2}^2=3\sigma_{1,1}^2$ gives $\alpha_1=3/4$ and $\alpha_2=1/4$. This reflects coefficients are inversely proportional to the channel noise contribution.

\subsection{Noise decorrelation}
\label{sec:noise_decorrelation}
The information on the correlated noise between channels can be useful to analyze the possible sources of noise in the system and also, for some applications, can be used to further reduce the noise in the output images. The noise reduction techniques by suppressing correlated noise could provide a way to improve the noise performance in applications with limited access to hardware modifications \cite{Rauscher_NSClean}. Typically, the largest noise correlation between channels happens at the same time in all the amplifiers, while in the MAS-CCD, the charge of a pixel is measured at different times (since the charge has to be moved in the serial register for that). If one amplifier is reading a charged pixel, while all of the others or a few of them are reading empty pixels, the noise information from those can be used to remove the common noise contribution from the first amplifier.

The new pixel value, denoted as $\hat{s}_i$, is the result of subtracting equalized empty pixels $\tilde{s}_k$ (with $k\neq i$) from the original one $\tilde{s}_i$. It is important to emphasize that for this algorithm, $\tilde{s}_i$ and $\tilde{s}_k$ are measurements of different charge packets that occur in the $N_a=16$ amplifiers at the same time, because of this, although $\tilde{s}_i$ could be a charged pixel, the other amplifiers measurements $\tilde{s}_k$ could be empty pixels. For those pixels $\tilde{s}_k$ that are eventually charged, a null weight is assigned. This can be formally expressed as
\begin{equation}
\hat{s}_i = 
\begin{cases} 
\tilde{s}_i + \sum\limits_{k=2}^{N_a} \alpha_{i,k} \tilde{s}_k, & i=1, \\
\tilde{s}_i + \sum\limits_{k=1, k \neq i}^{N_a} \alpha_{i,k} \tilde{s}_k, & 1<i<N_a, \\
\tilde{s}_i + \sum\limits_{k=1}^{N_a-1} \alpha_{i,k} \tilde{s}_k, & i=N_a.
\end{cases}
\label{eq:decorrelacion}
\end{equation}

The weights $\alpha_{i,k}$ involved in the previous expression are obtained similarly to those in Sec. \ref{Sec:optimizedAverge}, by performing minimization of the variance $\var(\hat{s}_i)$, though without the need to impose a restriction. They are determined by solving the subsequent linear system of equations for each of the $1\leq i\leq N_a$ channels as
\begin{equation}
    \begin{bmatrix}
    \sigma_1^2 & \cdots & \sigma_{1,k}^2 \\
    \vdots & \ddots & \vdots \\
    \sigma_{k,1}^2 & \cdots & \sigma_k^2
    \end{bmatrix}
    \begin{bmatrix}
    \alpha_{i,1} \\
    \vdots \\
    \alpha_{i,k}
    \end{bmatrix}
    =
    \begin{bmatrix}
    \sigma_{i,1}^2 \\
    \vdots \\
    \sigma_{i,k}^2
    \end{bmatrix}
    \label{eq:decorrelation_weights}
\end{equation}
where if $i=1$, every element starting at $1$ is instead considered starting at $2$. The square matrix represents the covariance matrix of the channels containing empty pixels, and the column vector on the right-hand side represents the covariances of the $i$-th channel with respect to the rest of the channels. This technique is applied before the optimum average of the channels, presented in Sec.\ref{Sec:optimizedAverge}.

A potential use of this type of technique is in spectroscopy where projected spectral lines appear separated by a few empty pixels on the CCD\cite{Guy_2023}. If the intermediate pixels have a low background contribution, their pixel value can be used to decorrelate the noise from channels measuring actual spectral lines. This could add a link between the length of pixel separation between amplifiers and the space between projected spectral lines in the active region.

\section{Node removal efficiency (NRE) of the sense nodes charge packets}
\label{sec:gre}

In this section, we present a model of the process of charge transfer between the amplifiers in the output stage. As this configuration of inline amplifiers is the main difference of the MAS-CCD compared to the Skipper-CCD it is important to characterize it in terms of noise and also in terms of its efficiency (or inefficiency) for transferring the charge packets from one amplifier to the next one.

For the model, we assume that a fraction of the charge is left behind in the sense node of each amplifier when the packet is taken out to be transferred to the next amplifier. To differentiate this phenomenon from the standard CTI/CTE process in CCDs, the efficiency for removing the charge from the sense nodes is defined as node removal efficiency (NRE) and node removal inefficiency (NRI). This mechanism can be explained using Fig.~\ref{fig:output_stage}\textcolor{blue}{b}. Once the charge in the channel of the serial register is measured in each output stage by the amplifiers (A$_i$) through the non-destructive sense nodes (SN), the charge should be moved out using the Pixel Separation gate (PS). The PS voltage is moved with H$_{1}$. Then, the typical three-phase clock sequence allocates the charge under H$_2$ before the next pixel readout. With this sequence, the charge is removed from the SN and incorporated into the next serial register pixels. During the charge removal from the SN, some of the carriers could be left behind. A fraction of the measured charge by one amplifier could stay in the sensor and be added to the following pixel charge carrier. One of the critical aspects is that this extra charge does not affect the measured value of the next pixel since it will be part of the reference voltage for the DSI calculation (the pedestal level), however, this fraction of extra charge will affect the next amplifier. Looking at the output images, one of the characteristic signatures of this effect is that the first amplifier in the chain does not see the effect of the NRE, even in a situation where the effect is very aggressive in the other channels.

\begin{figure}[h!]
    \centering
    \includegraphics[width=1\linewidth]{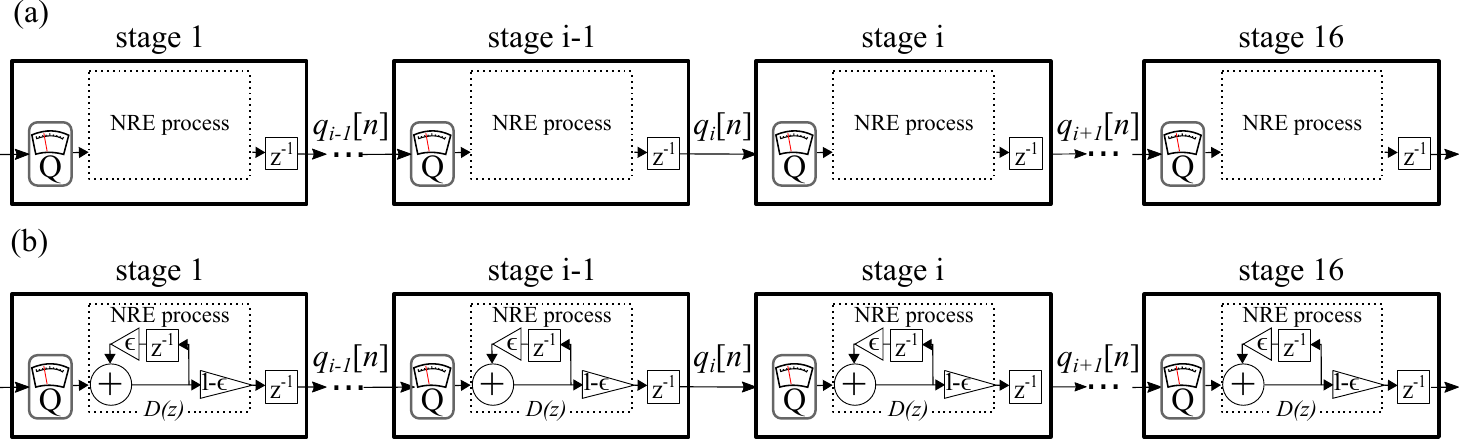}
    \caption{Model of the charge transfer among the different output stages. In this model, we are only considering the transfer inefficiency introduced by the new inline amplifiers, and we are not modeling charge transfer inefficiencies in the pixels connecting the stages.}
    \label{fig:gre_model}
\end{figure}

The NRE process can be thought as depicted in the block diagram of Fig.~\ref{fig:gre_model}\textcolor{blue}{a}. The first amplifier (stage 1) measures the charge without NRE issues, illustrated by the meter before the NRE process. After that the NRE process takes place, distorting the charge packet. The next amplifier stage will measure this effect in the next time instant (represented by the ideal delay block $z^{-1}$).

\subsection{NRE between two consecutive amplifiers}

The ``NRE process'' block in Fig.~\ref{fig:gre_model}\textcolor{blue}{a} can be modeled as a recursive discrete system. Following the methodology in \cite[Chapter~3, pp.103-108]{howes1979charge} we write the equation for the value of the next charge packet $q_i[n+1]$ (time instant $n+1$) measured in amplifier $i$ as a function of the previous charge packet $q_{i}[n]$ (time instant $n$) measured in the same amplifier and the charge packet measured in the predecessor amplifier $q_{i-1}[n]$. 
\begin{align}
    q_i[n+1]&=(1-\epsilon)\left(q_{i-1}[n]+\frac{\epsilon}{(1-\epsilon)} q_{i}[n]\right) \nonumber \\
    &=(1-\epsilon)q_{i-1}[n]+\epsilon q_{i}[n]
    \label{eq:Recursive_qj}
\end{align}
where $(1-\epsilon)$ is the NRE and $0\leq \epsilon \leq 1$ the NRI. 

The first line in Eq.(\ref{eq:Recursive_qj}) represents the efficiency $1-\epsilon$ of transferring the charge packet that was in the previous amplifier $i-1$ sense node to the next amplifier $i$, with $i\geq 2$. This charge packet has two components: 1) the part that was effectively measured by the previous amplifier $q_{i-1}[n]$ and 2) the fraction of the charge that was present in the SN of amplifier $i-1$ but was not measured because it was part of the reference voltage (pedestal level). This fraction of charge was lost from the charge packet that now is in the sense node of amplifier $i$ and can be estimated from the measurement $q_{i}[n]$ as $\frac{\epsilon}{(1-\epsilon)} q_{i}[n]$, where $q_{i}[n]/(1-\epsilon)$ estimates the charge packet when it was in the previous amplifier and that times the inefficiency $\epsilon$ estimates the fraction of charge in point 2). 

It is worth noting that solving Eq.(\ref{eq:Recursive_qj}) for the NRI results in 
\begin{equation}
\epsilon=(q_{i}[n]-q_{i-1}[n-1])/(q_{i}[n-1]-q_{i-1}[n-1]), \nonumber
\end{equation}
which gives a formula for computing the inefficiency given the measurements of two consecutive amplifiers.

Equation (\ref{eq:Recursive_qj}) is a recursive difference equation that can be solved to get the charge measured in amplifier $i$, $q_{i}$, as a function, only, of the charge in the previous amplifier $q_{i-1}$. Using the $z$-transform for discrete dynamic systems:
\begin{equation}
Q_i(z)=D(z)z^{-1} Q_{i-1}(z),
\label{eq:TZOneDelay}
\end{equation}
where the transfer function 
\begin{equation}
    D(z)=\frac{(1-\epsilon)}{(1-\epsilon z^{-1})}
    \label{eq:Dz}
\end{equation}
models the distortion effect of the NRE process on the charge packet measurement between two consecutive amplifiers. 

A block diagram of this process is shown in Fig.~\ref{fig:gre_model}\textcolor{blue}{b}, where for each stage, the pixel value is obtained first, the NRE process is modeled by the distortion transfer function $D(z)$ and the pure delay $z^{-1}$ models the deferred measurement between two consecutive inline amplifiers.
In other words, if the NRE was perfect ($\epsilon=0$) then $Q_i(z)=z^{-1} Q_{i-1}(z)$ and since $z^{-1}$ is just a pure delay, anti-transforming results in $q_i[n]=q_{i-1}[n-1]$, i.e., the charge in amplifier $i$ in time instant $n$ is exactly equal to the charge measured in the previous amplifier $i-1$ in the previous time instant $n-1$.

\subsection{NRE in the $N_a$ inline amplifiers}
From the point of view of the NRE, the distortion effect in the $N_a$ inline amplifiers, can be seen as a cascade connection of the $D(z)$ transfer functions. An important point to make is that the first amplifier measures the charge packet without NRI related issues because in the first sense node no charge was lost yet. Therefore, this measurement can be considered as the ideal input signal to the rest of $N_a-1$ inline amplifiers.
To model the effect of the NRE at any of the amplifiers, after the first one, the cascading of distortion transfer functions results in
\begin{equation}
    D_{n_a}(z)=\frac{(1-\epsilon)^{n_a-1}}{(1-\epsilon z^{-1})^{n_a-1}},
    \label{eq:Dz_na}
\end{equation}
with $2\leq n_a \leq N_a$. This transfer function can be used to predict the effect of the NRE for a given stream of input charge packets. 
A well-known input could be obtained by applying a flat field of light to the sensor followed by a readout with the $N_a$ amplifiers that extend beyond the number of columns of the active region, producing overscan pixels that should be empty if NRE is perfect. This is similar to the extended pixel edge response (EPER) method for CTI \cite[Chapter~5, pp.423-429]{janesick2001scientific} and results in a negative step-function input of charge for the $N_a$ inline amplifiers. 

Assuming a negative step function of $Q$ electrons (i.e., the charge decreases from 0 to $-Q$) with z-transform $-Q/(1-z^{-1})$, as input to $D_{n_a}(z)$, the output in the z-domain is given by $Y_{-Q}(z)=-Q D_{n_a}(z) /(1-z^{-1})$. The time-domain output $y_{n_a}[n]$ for the $n_a$-th amplifier, caused by the NRE due to a negative step of charge (from $Q$ to 0 electrons), is $y_{n_a}[n]=\mathcal{Z}^{-1}\{Y_{-Q}(z)\}+Q$ for $n\geq0$. The anti-transform is performed by partial fraction expansion for the pole of order $n_a-1$ in (\ref{eq:Dz_na}), obtaining:
\begin{equation}
y_{n_a}[n]=Q\epsilon^{n+1} u[n]\sum_{m=1}^{n_a-1} \frac{(1-\epsilon)^{m-1}\prod_{u=1}^{m-1}(n+u)}{(m-1)!} 
\label{eq:ynan}
\end{equation}
where $u[n]$ is the unit step function, $2\leq n_a \leq N_a$ and $n \geq 0$. Equation (\ref{eq:ynan}) predicts the output in the overscan pixels ($n=0$ is the first overscan pixel) for each of the amplifiers.

For example, for $n_a=2$ the charge measured by the second amplifier after a perfect step of charge measured by the first amplifier is:
\begin{equation}
y_{2}[n]=Q\epsilon^{n+1} u[n],
\end{equation}
which is an exponential response. Usually, for CTI measurements with EPER method, only the first overscan pixel is measured. For this amplifier this results in $y_{2}[0]=Q\epsilon$, which as expected corresponds to the original charge packet $Q$ times the NRI, the inefficiency for removing the charge. 

The first over-scan pixel in each of the amplifiers after the first amplifier is 
\begin{equation}
y_{n_a}[0]=Q-Q (1-\epsilon )^{n_a-1}
\label{eq:yna0FirstOSPix}
\end{equation}
with $2\leq n_a \leq N_a$. For $\epsilon\ll 1$ this can be approximated by a straight line resulting in
\begin{equation}
\label{eq:gre proportional aproximation}
   y_{n_a}[0]\approx -Q \ln(1-\epsilon) (n_a-1)\approx Q \epsilon (n_a-1). 
\end{equation}
\subsection{Average NRE/NRI of the amplifier chain}
In the previous description, $\epsilon$ represents the individual NRI for each amplifier. In the MAS-CCD presented in this work, the final pixel value is obtained by averaging the measurement in the $N_a$ amplifiers. Therefore, the real impact on the final pixel value of the node removal efficiency modeled in this section should be computed as the average of the outputs in each of the amplifiers. Assuming a relatively low $\epsilon$ most of the impact of the NRI is present in the first overscan pixel $y_{n_a}[0]$, using Eq.(\ref{eq:yna0FirstOSPix}), the average NRI for the amplifier chain can be computed by averaging $y_{n_a}[0]$ for $2\leq n_a \leq N_a$ and dividing by the total charge $Q$

\begin{equation}
\epsilon_{avg}=\frac{N_a^{-1}}{Q}\sum_{n_a=2}^{N_a}y_{n_a}[0]=\frac{(1-\epsilon)^{N_a}+N_a \epsilon-1}{N_a \epsilon} \approx \frac{N_a-1}{2} \epsilon,
\end{equation}
where for the approximation it is assumed that $\epsilon\ll 1$.

To put this in context, it is worth comparing the signal degradation by NRE (or NRI) to the one from the CTI in the horizontal register. Considering a detector of N$_{col}$ columns and a small CTI and NRI so its effect would be proportional to the number of transfers, the total charge left behind from a pixel with $Q$ carriers is $Q\times CTI\times N_{col}$ for the pixel transfer, and $Q\times \epsilon_{avg}\times N_{a}$ due to the node removal inefficiency. Then, the NRI that will produce the same degradation that a given CTI is $\epsilon_{avg} = CTI\times N_{col}/N_{a}$. For example, assuming a sensor with $N_a=16$ inline amplifiers and N$_{col}=2000$ columns with a CTI = $1\times10^{-6}$, then $\epsilon_{avg} = 1.25\times10^{-4}$ to get a similar degradation of the charge signal. This example shows that CTI and NRI impact is different.

\subsection{Non-constant NRE/NRI}
It is well known that CTI/CTE has a dependence on the charge packet size\cite{villalpando_SOAR2024} and also on other factors such as temperature \cite{murowinski1995charge}. It is expected that the same nonlinear behavior could be present in NRE/NRI measurements. In this scenario, the recursive Eq.(\ref{eq:Recursive_qj}) is still valid, but the Z-transform analysis cannot be used. Nevertheless, an algorithmic approach using the recursion could be implemented. Appendix \ref{appendix_GRE} shows a complete and fully functional example of the recursive algorithm implemented for the nonlinear case.

\section{Experimental results}
\label{exp_results}

\begin{figure}[h!]
    \centering
    \includegraphics[angle=0,width=0.8\linewidth]{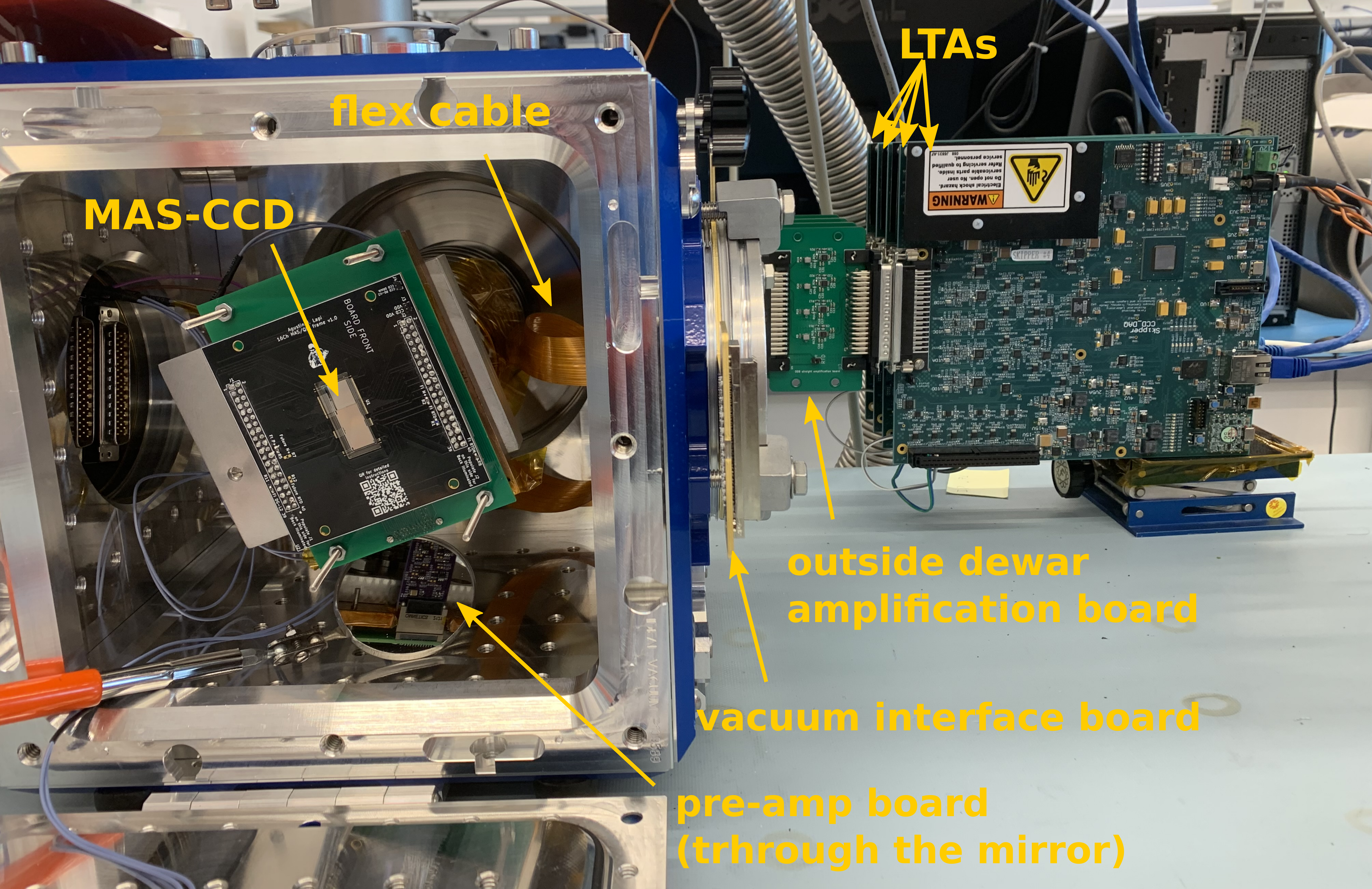}
    \caption{Experimental setup used for the characterization of the sensor.}
    \label{fig:setup}
\end{figure}

Figure \ref{fig:setup} shows the experimental setup used to characterize the proposed sensor. The sensor is assembled in a custom picture frame package (black PCB board) and installed inside a vacuum chamber for cooling purposes. Inside the dewar, the sensor signals traverse through a compact board with JFET amplifier stages, buffering the signals before being channeled through a flex cable to a vacuum interface board. The external connector links to a multi-readout controller bus, subsequently connecting to four Low Threshold Acquisition controllers (LTA \cite{cancelo2021low}) with four video channels each. Between the bus and the LTA, an external dewar amplification board provides additional amplification and buffering for the video signals. The four readout controllers operate synchronously in a leader-follower configuration. The leader board supplies clocks and biases for the sensor, while all controllers collectively process the output video signals.

Under the auspices of the U.S. Department Of Energy (DOE) Quantum Information Science initiative, $p$-channel 675$\mu$m-thick MAS-CCDs were fabricated on high-resistivity $n$-type silicon (${\sim}\,10$ k$\Omega$\,cm). The sensors were designed at LBNL to be operated as thick fully-depleted devices with high quantum efficiency over a broad wavelength range \cite{Holland:2003} (see \cite{holland_2023} for more details on the fabricated sensors). The sensors were fabricated at Teledyne DALSA Semiconductor, diced at LBNL, and packaged/tested at Fermilab.

\begin{figure}[h!]
    \centering
    \includegraphics[angle=0,width=1\linewidth]{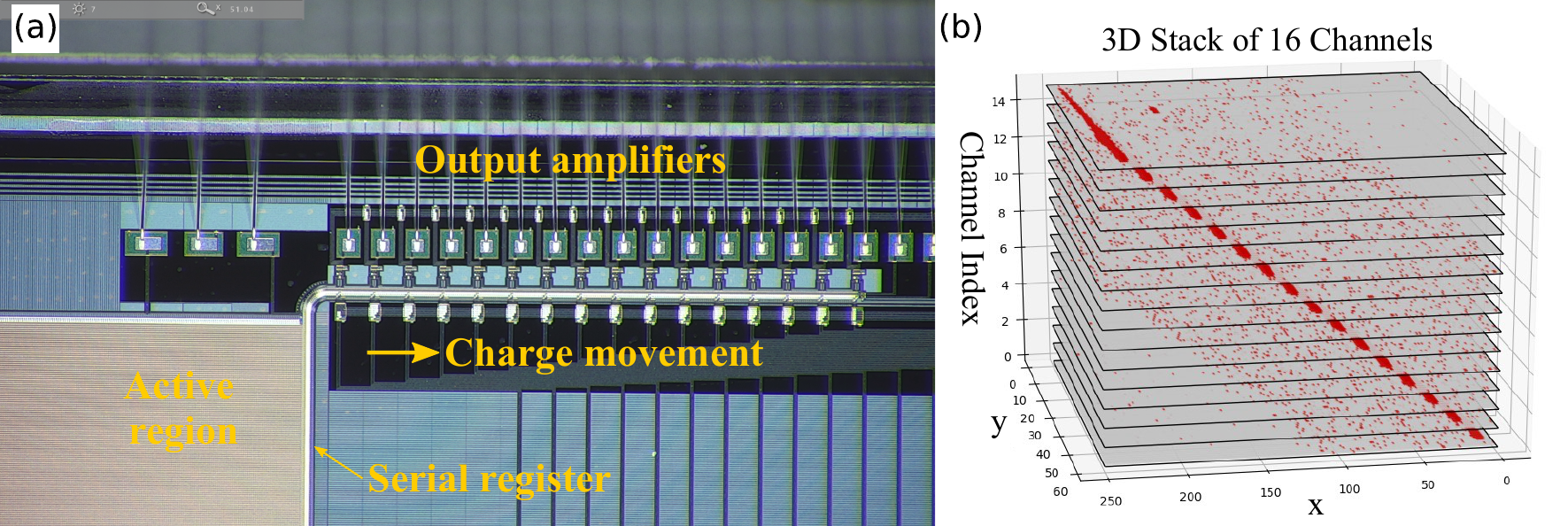}
    \caption{(a) Microscope picture of the MAS-CCD output stage with the 16 output amplifiers in the serial register. (b) A 3D representation of the output images from the transistors in normal operation.}
    \label{fig:micro_stack}
\end{figure}

The sensor under study possesses $N_a = 16$ non-destructive readout stages on the same serial line and operates by collecting holes. A microscope image of the output stage of the MAS-CCD is shown in Fig.~\ref{fig:micro_stack}\textcolor{blue}{a}. The figure displays the direction of charge movement through the multiple inline amplifiers. The active region is located in the bottom left corner, with the serial register indicated. The sensor consists of a 1024 $\times$ 512 array of 15\,$\mu$m $\times$ 15\,$\mu$m pixels and a substrate thickness of 675\,$\mu$m. The output amplifiers are separated by $15$ pixels. The pixels in the prescan are used to bend the serial register by 90 degrees as shown in the picture. The sensor is operated at 140K using a polycold cryocooler and in fully depletion mode using a substrate voltage of 70V.

Figure ~\ref{fig:micro_stack}\textcolor{blue}{b} displays a three-dimensional representation of the output images of the detector after its readout. The images are superimposed to illustrate the fixed separation (in number of pixels) between each output stage. An ionization event from natural radiation can be observed moving to the right as the images are scrolled from top to bottom. Color binarization was applied to enhance visual inspection. The results shown here were taken with the first packaged detector sixteen-channel MAS-CCD with all amplifiers working and no evidence of columns or bright defects were observed in the active area.

Each LTA performs a digital DSI to calculate the final pixel value which is sent through Ethernet to a computer. Offline software in the computer is used to process the output data to get the final image product.

\subsection{Readout noise studies}

\begin{figure}[t!]
  \begin{subfigure}{\textwidth}
    \centering
    \caption{}
    \includegraphics[trim=0 0 0 8mm,clip,width=0.8\linewidth]{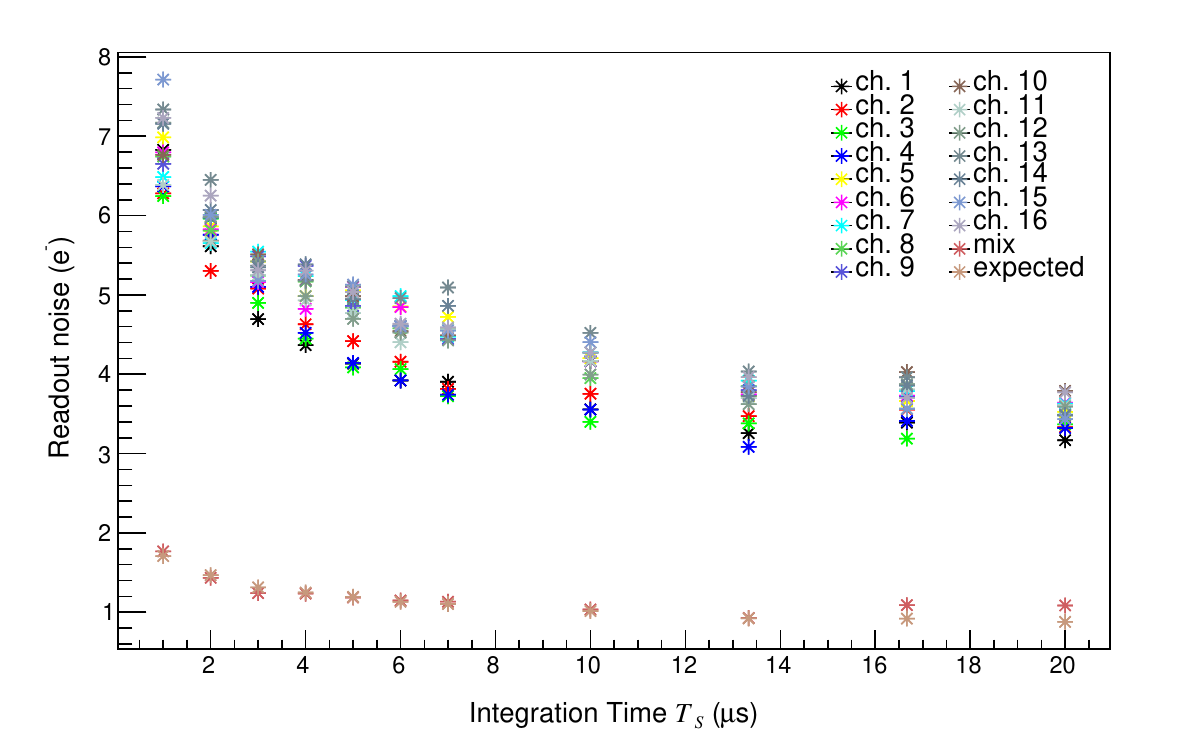}
     \label{fig:xsamp_a}
  \end{subfigure}
  
  \begin{subfigure}{\textwidth}
    \centering
    \caption{}
    \includegraphics[trim=0 0 0 8mm,clip,width=0.8\linewidth]{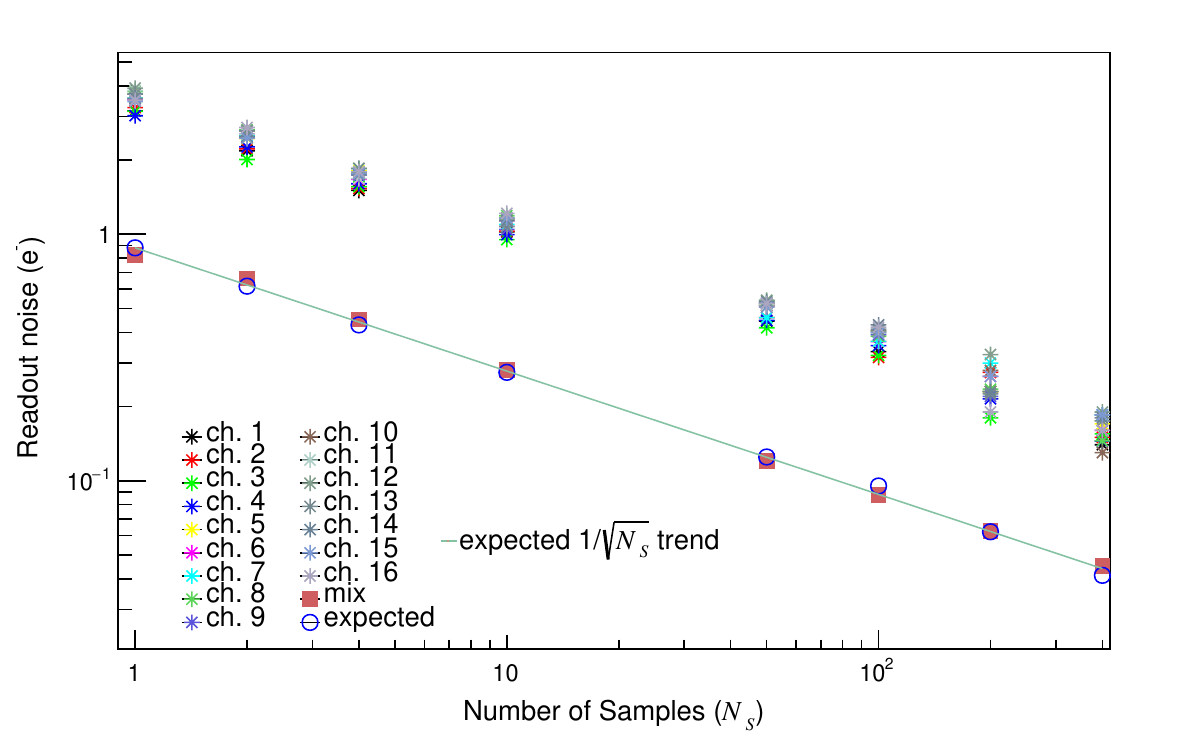}
     \label{fig:xsamp_b}
  \end{subfigure}
  
  \caption{a) Noise performance for different Integration times $T_s$ of all the channels, and combination and expected performance. b) Noise reduction of all the channels as a function of the Skipping samples $N_s$ and the standard and expected combination of the channels, for an integration time $T_s=16.66\mu$s.}
  \label{fig:xsamp}
\end{figure}

This section demonstrates the noise reduction properties of the sensor operated in different conditions of integration time and the number of samples $N_s$ taken by each amplifier. Figure \ref{fig:xsamp_a} shows the readout noise performance of the 16 amplifiers as a function of the readout integration time ($T_s$) of the DSI operation (mathematical model presented in Appendix \ref{appendix_freq_response}), using $N_s$ = 1 per amplifier. Similar operation is obtained from all of them. At the same time, the plot shows the measured noise after combining the information from all the channels with equal weights, following Eq.(\ref{eq:average equal weights}). The plot also shows the expected noise performance after this calculation which is given by Eq.(\ref{eq:expected noise}), both quantities have a good agreement.

The noise performance was also measured as a function of the number of samples per amplifier ($N_s$) using an integration time of $T_s$ = 16.6 $\mu$s is shown in Fig.~\ref{fig:xsamp_b}. All the amplifiers show similar noise performance. The plot also shows in circled markers the resultant noise when the information from the channels is averaged using Eq. (\ref{eq:average equal weights}). The combined measurement shows good agreement with the expected performance (square markers) from Eq. (\ref{eq:expected noise}). The expected combined noise value for $N_s=1$ is used to extrapolate to larger $N_s$ dividing by $\sqrt{N_s}$ giving a straight line in the plot, which is indicated as a green line in Fig.\ref{fig:xsamp_b}. The line reveals that the measured performance follows the expected reduction fraction.

\subsection{Performance at targeted speeds and implementation of processing techniques}
\label{sec:fast_readout}

This section focuses on the study of the noise performance using the noise reduction properties at specific target readout speeds and includes the optimization techniques developed in Sec. \ref{data_procesin}.

One important goal of this work is to demonstrate that the technology can achieve low noise levels while running at a readout speed similar to the one used in the Dark Energy Spectroscopic Instrument (DESI) \cite{Bebek_2017, Abareshi_2022} and to use the developed techniques to study and optimize the response of the sensor. 

\begin{figure}[t!]
    \centering
    \includegraphics[angle=0,width=0.85\linewidth]{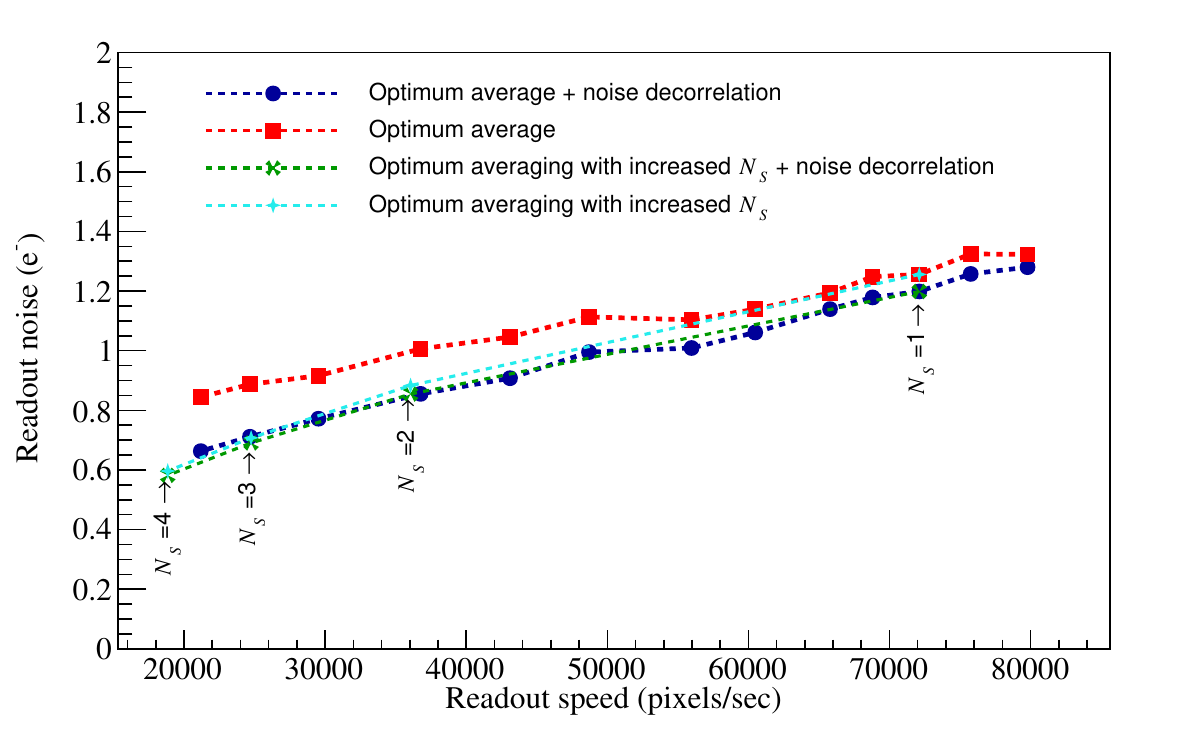}
    \caption{Measured noise performance as a function of the pixel readout speed. Blue circles and red squares with and without noise decorrelation, respectively, for $N_s=1$. Also, in green and cyan markers, with and without noise decorrelation, for $N_s=2, 3,$ and $4$.}
    \label{fig:noise vs speed}
\end{figure}

Figure \ref{fig:noise vs speed} shows the noise performance as a function of the pixel readout speed, in terms of pixels per second. 
The plot presents the performance applying the different methodologies explained in the previous sections. The red square represents the noise performance using one sample per amplifier ($N_s=1$) and equalization and an optimum average of the channels from Sec. \ref{sec:gain_equalization} and \ref{Sec:optimizedAverge}, respectively. The gain equalization of the channel was performed using ionizing events from the natural radioactivity. The readout clock sequence was developed following the recipe provided in \cite{lapi_fast} for fast readout of Skipper-CCD sensors. The different speeds are obtained by increasing the integration time in the charge measurements. There was no individual optimization of the clock sequence and voltages for different speed points. The measurements show around 1 e$^-$ of noise performance for the evaluated speeds.

\begin{figure}[b!]
    \centering
    \includegraphics[angle=0,width=1\linewidth]{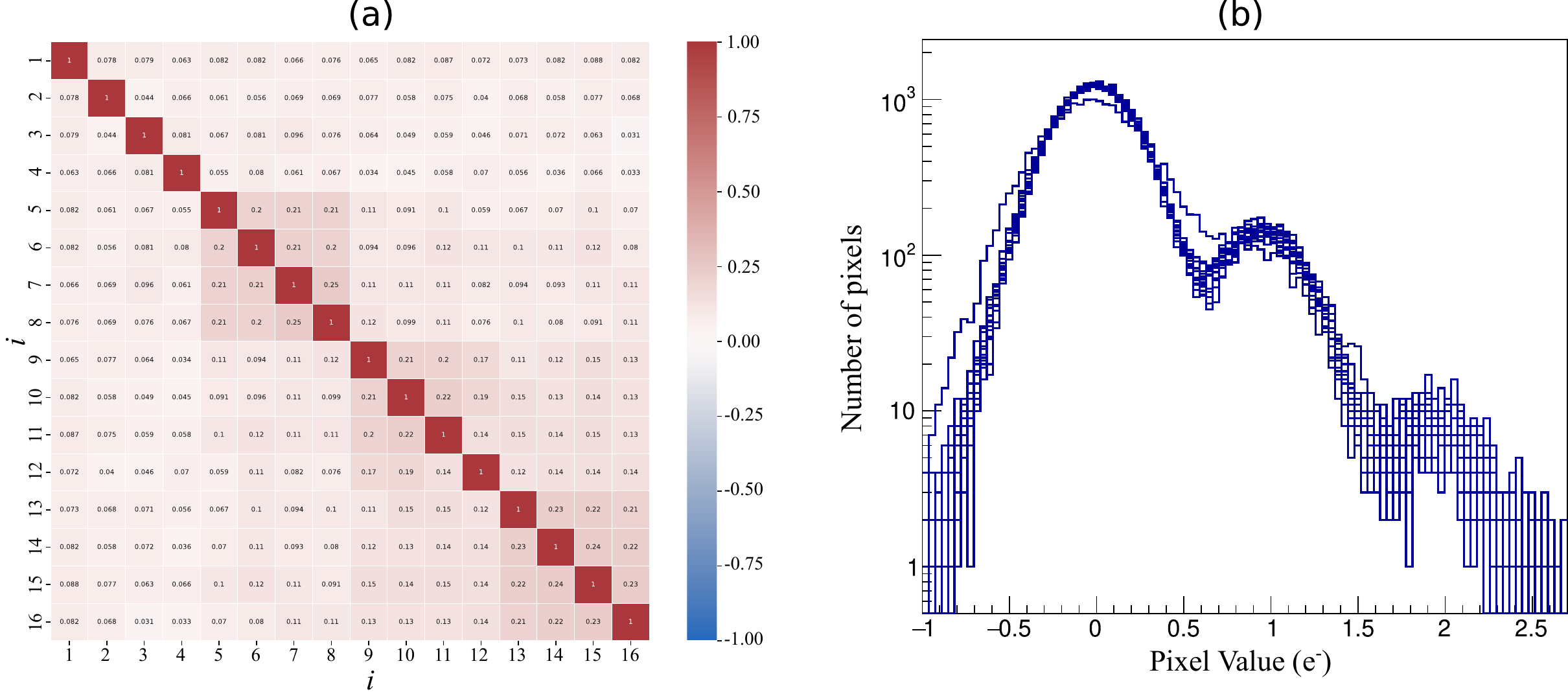}
    \caption{(a) Measured noise correlation matrix with the detector running at $~72\times10^{3}$ pixels/sec. (b) Histogram of the pixels of all the amplifiers when the sensor is running at a $~72\times10^{3}$ pixels/sec using 400 samples per amplifier.}
    \label{fig:correlation_peaks}
\end{figure}

Using the noise decorrelation technique from Sec. \ref{sec:noise_decorrelation} applied to the same collected data, the obtained performance is shown with the blue circles in Fig.~\ref{fig:noise vs speed}. This process helps to reduce the correlated noise of the channels using the noise information from other channels. Although the effect seems to be small at the faster readout, its contribution gets to around 25\% at the slowest evaluated readouts. To diagnose the common noise origin we evaluated the covariance matrix from Eq.(\ref{eq:covariance matrix}), which is shown in Fig.~\ref{fig:correlation_peaks}\textcolor{blue}{a}. The color scale shows the correlation index. There seems to be a larger correlation in groups of four amplifiers, which correspond to the channels that are read out by the same LTA. This could be related to grounding issues in the interconnection of multiple LTAs. We expect to fix this problem in future measurements.

Another way to read the sensor is to allow each amplifier to take several samples. The light blue plus markers in Fig.~\ref{fig:noise vs speed} show the noise performance of the system when multiple samples are taken on each amplifier. To compute the final pixel value, the first step is to average the samples from individual amplifiers using weights of 1/$N_s$. Secondly, the resulting values from the amplifiers are optimally combined following the recipes in Sec. \ref{sec:gain_equalization} and \ref{Sec:optimizedAverge}. These measurements were performed by increasing the number of samples per amplifier from the original readout speed of $\sim$72k pixels/sec, indicated in the plot with the label $N_s=1$. The dashed lines of the same color visually connect these points in the plot. It is interesting to note that increasing $N_s$ gives a better noise performance than the red square dots, which represent images processed with the same techniques. It is expected that the performance of the multiple skipping samples for the same total pixel time could give a different noise performance due to the different effective frequency responses. Please see the appendix \ref{appendix_freq_response} for a discussion about the expected frequency of the system under different parameter conditions. One interesting feature to note is that the measured noise performance from multiple samples follows the expected reduction by the factor $\sqrt{N_s}$ from the starting point at $N_s=1$.
Applying the noise decorrelation technique (discussed in Sec. \ref{sec:noise_decorrelation}) on the multiple $N_s$ measurements we obtained the performance shown with cross markers in green.  
This processing has little effect on the performance indicating that there is little contribution of common noise to subtract between amplifiers. The multiple sampling operation per amplifier provides an extra degree of freedom for the optimization of the readout operation of this kind of sensor. If the multiple sampling readout of the entire pixel is a limitation for the measurements of all the pixels in the active regions, a selective strategy with regions of interest as in \cite{chierchie2020smartreadout} could be used to increase the signal-to-noise ratio of interesting pixels.

When $N_s$ is further increased the noise can be reduced to arbitrary levels. Fig.~\ref{fig:correlation_peaks}\textcolor{blue}{b} presents $N_a$ superimposed histograms, one for each amplifier. The peaks represent pixels with a discretized amount of collected carriers. This is helpful for absolute calibration of the gain of the individual channels \cite{Rodrigues_2020} and in applications with single photon requirements as in the case of exoplanet searches. Although peaks are observed for all the channels, channel 13 shows wider distributions.

\subsection{Node removal efficiency}

\begin{figure}[h!]
    \centering
    \includegraphics[angle=0,width=1\linewidth]{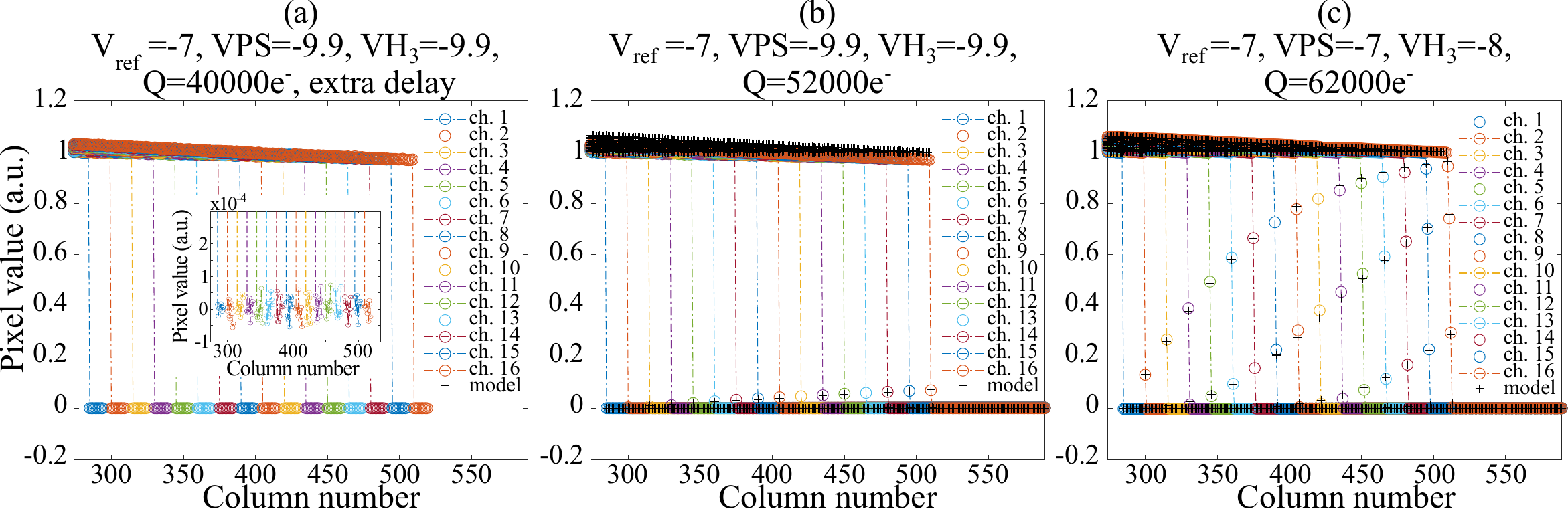}
    \caption{Measured charge value as a function of the image's column index for the sixteen MAS-CCD channels. In particular, the plot shows the transition between the pixel in the active region and the overscan region. The different plots show different levels of NRE effect on the measured signal.}
    \label{fig:gre measurements}
\end{figure}

As discussed in Sec.~\ref{sec:gre}, the NRE has a strong dependency on the operation of the sensor: voltages, size of the charge packet, and the speed of the clocks. The goal of this section is to study experimentally the NRE for different cases of $\epsilon$ and provide ideas on how to operate this sensor compared to the regular single-stage detectors. For this task, we show three witness cases which are represented by the plots in Fig.~\ref{fig:gre measurements}. The plots show the measured pixel value as a function of the column index. Each trace is the average of 15 rows that are averaged to reduce the uncertainty in the pixel measurements. The plots show the transition from the active to the overscan pixels for all the channels of the sensor, producing a negative step of charge. The measured pixels are shown in circle markers and a dashed line with the same color is used to connect them and to help the reader to follow the abrupt transition between the regions. The sensor was illuminated prior to the readout and areas with uniform light exposure were used. A moderate amount of charge was intentionally used in the pixels to show the behavior in a realistic scenario. The black dots in some of the plots represent the fit of the proposed NRI model to the data. Each plot has a title identifying the most important conditions of voltages in the gates of the sense node (V$_{ref}$) and the low levels of the PS (VPS) and H (VH) clocks, as well as the amount of charge used in each case. To generate the plots, the offset of each image is subtracted and the values of the pixels are normalized by the amount of charge in the active pixels a few pixels before the transition to the overscan. This representation helps to easily infer the $\epsilon$ values from the plots using the normalized value of the pixels in the overscan. The following scenarios were considered:

\begin{itemize}
    \item NRI below the precision of the method. In Fig.~\ref{fig:gre measurements}\textcolor{blue}{a}, the first pixel in the overscan (i.e., the first pixel after the abrupt transition) does not have an evident trace of charge for all the channels. The inlet shows a zoom-in in the overscan regions. The first pixel seems to have a statistic similar to the rest of the pixels of the overscan. In this case, $\epsilon < 5\times 10^{-5}$. 

    \item Moderate NRI. Figure \ref{fig:gre measurements}\textcolor{blue}{b} shows the case where the first pixel in the overscan starts growing as a function of the number of amplifiers proportionally, such as in Eq.(\ref{eq:gre proportional aproximation}). In this case, $\epsilon=5\times10^{-3}$ from the fitted model, which is shown with the black dots.

    \item Large NRI. Figure \ref{fig:gre measurements}\textcolor{blue}{c} shows an extreme case where the PS and VH low-level voltages of the clocks are set close to the gate voltage of the sense node to force a bad charge removal from the sense node (very large NRI). The intermediate pixels in the transitions from the active to the overscan regions are due to this poor removal capability. In this case, the NRI is very severe, and the pixels in the overscan do not show a proportional growth with the amplifier number. In this extreme case, the best-fit model was obtained by assuming a $\epsilon$ with proportional dependence $\epsilon = 2.26\times10^{-6}Q + 3\times10^{-4}$. The algorithmic model, presented in Appendix \ref{appendix_GRE}, was used for the calculation.
    
\end{itemize}

It is interesting to note that even though the removal inefficiency is very large, the first pixel of the overscan in the first amplifier does not show any sign of this problem, similar to the other plots explained before. This proves the hypothesis that this is a new type of charge inefficiency linked to the process of moving out the charge of one amplifier after its measurement. 

Another important aspect to note is that the removal of the charge from the sense nodes requires voltages below the sense node reference voltage. For the case of CCD collecting holes, this means that the voltages should be more negative than the sense node gate voltage. The current limitation of our hardware does not allow us to explore clock voltages below -10 volts; there is an opportunity for future measurements to improve the NRE by using smaller clock values. Another aspect to incorporate in future instrumentation of this kind of sensor is the use of a clock to set voltages in the sense node instead of a fixed bias voltage (as currently being used). A clock would allow an increase in the voltage in the sense node when transferring the change to increase the gradient towards the PS gate.

\section{Conclusions}
\label{sec:conclusion}
In this work, a 16-in-line MAS-CCD was presented as a promising sensor candidate for the next generation of astronomical instruments, such as the Spec-S5. At the same time, a new theoretical framework was developed to address the unique properties of this type of detector for future optimization and characterization. Among these techniques, an optimum noise averaging method, a common noise subtraction technique, and a model of charge removal inefficiencies from the output stages were presented. Moreover, these introduced processing methods and theoretical frameworks were supported by relevant experimental results.

\section* {Code, Data, and Materials Availability}
The code to compute the NRE (Sec. \ref{sec:gre}) is presented in the Appendix \ref{appendix_GRE}. The data that supports some of the findings of this article are not publicly available due to the large size of datasets and the individual files and privacy concerns. They can be requested from the corresponding author's email address.

\section* {Acknowledgments}

The fully depleted Skipper-CCD was developed at Lawrence Berkeley National Laboratory, as were the designs described in this work. The CCD development work was supported in part by the Director, Office of Science, of the U.S. Department of Energy under No. DE-AC02-05CH11231.
The multi-amplifier sensing (MAS) CCD was developed as a collaborative endeavor between Lawrence Berkeley National Laboratory and Fermi National Accelerator Laboratory. Funding for the design and fabrication of the MAS device described in this work came from a combination of sources including the DOE Quantum Information Science (QIS) initiative, the DOE Early Career Research Program, and the Laboratory Directed Research and Development Program at Fermi National Accelerator Laboratory under Contract No. DE-AC02-07CH11359.

This research has been partially supported by a Detector Research and Development New Initiatives seed grant at Fermilab, a grant from the Heising-Simons Foundation (\#2023-4611), and Javier Tiffenberg's and Guillermo Fernandez Moroni's DOE Early Career research programs.

\bibliography{main}
\bibliographystyle{spiejour}

\listoffigures

\appendix  
\section{Impulse and frequency response of the MAS-CCD to common noise}
\label{appendix_freq_response}

The Dual-Slope Integrator (DSI) impulse response and frequency response are \cite[Chapter~6, pp.574-581]{janesick2001scientific}\cite{skipper_2012}
\begin{equation}
 h_{DSI}\left( t\right) =\begin{cases}\dfrac{A}{T_{s}}, \hspace{0.2cm}
t_0 \leq t\leq t_0+T_{s}\\
-\dfrac{A}{T_{s}},
t_0+T_{s} \leq t\leq t_0+2T_{s} \\0, \hspace{1cm} \text{in other case}
\end{cases}
\label{eq:h_dsi}
\end{equation}
\begin{equation}
    H_{DSI}\left( f\right) = \frac{2A\sin^2\left(\pi T_sf\right)}{\pi T_sf} \nonumber
\end{equation}
where $T_s$ is the integration time for one sample and $t_0$ is an arbitrary start point of the integration. For the frequency response is assumed $t_0=0$ without loss of generality.

Extending the expression to calculate the impulse response of the multiple readouts of a single amplifier, results in the following impulse and frequency response for the Skipper readout \cite{cancelo2021low}
\begin{equation}
 {h_{SKP}\left( t\right) =\dfrac{1}{N_{s}}\sum ^{N_{s}-1}_{n=0}h_{DSI}\left( t-n\left( \tau+2T_{s}\right) \right)}
 \label{eq:h_skiper}
\end{equation}
\begin{equation}
   H_{SKP}\left( f\right) = \dfrac{H_{DSI}\left( f\right)}{N_s} \left| \frac{\sin\left(2\pi N_sT_sf\right)}{\sin\left(2\pi T_sf\right)}\right| \nonumber
\end{equation}
where $\tau$ is the time between consecutive samples, for simplicity $\tau = 0$. 

The transfer functions in (\ref{eq:h_dsi}) and (\ref{eq:h_skiper}) with a given PSD measurement of $n_{ai}$ for each amplifier can be used, to evaluate how $n_{ai}$ affects the pixel value. Integrating the output PSD over the frequency range of interest gives $\sigma_{i}$ of (\ref{eq:expected noise}).

An additional transfer function is needed if there is also a common mode noise source $n_{CM}$. Following this idea the, $h_{SKP}$ can be extended to calculate the response of the MAS-CCD readout to common noise

\begin{dmath}
\label{eq:MAS}
h_{MAS}\left( t\right) =\dfrac{1}{N_{a}}\dfrac{1}{N_{s}}\sum ^{N_{a}-1}_{m=0}\sum ^{N_{s}-1}_{n=0}h_{DSI}\left( t-m\Delta t -2nT_{s}\right)
\end{dmath}
where $\Delta t$ model the required transfer time and intermediate readout time in between stages.

In the continuous readout of the sensor, all the pixels are moved sequentially, and the time the charge spends between amplifiers is equal to the number of pixels in the serial register between consecutive amplifiers multiplied by the total readout time of one pixel which is $\Delta t$.

To compute the frequency response of the MAS sensor operation for a common mode noise input, the Fourier transform of the impulsive response in Eq.(\ref{eq:MAS}) has to be computed. Applying standard properties of the Fourier transform, building on the frequency response of the DSI and the Skipper operation with a single amplifier \cite{cancelo2021low}, the resultant frequency response of the MAS sensor can be found to be:
\begin{equation}
\left| H_{MAS}\left( f\right) \right| =\dfrac{2A\sin^2\left(\pi T_sf \right) }{\pi N_s T_s N_{a} f} \left| \dfrac{\sin\left( 2 \pi N_{s}T_{s}f\right) }{\sin\left( 2 \pi T_{s}f\right) } \dfrac{\sin\left[ \pi N_a \left(\Delta t+2N_sT_s\right) f \right] }{\sin\left[ \pi \left(\Delta t+2N_sT_s\right) f \right] }\right|
\label{eq:H_MAS_response}
\end{equation}

With this transfer function, and with the power spectral density (PSD) of the common mode noise, the PSD at the output of Eq.(\ref{eq:H_MAS_response}) can be calculated and integrate the noise in the operating frequency range, to obtain the variance $\sigma^2_{n_{CM}}$ of the common mode noise at the output of the MAS. The total noise is the quadrature sum of these two noise contributions, being 
\begin{equation}
\sigma_{tot} = \sqrt{\sigma^2 + \sigma^2_{n_{CM}}}    
\end{equation}
\begin{figure}[t]
    \centering
    \includegraphics[width=0.7\linewidth]{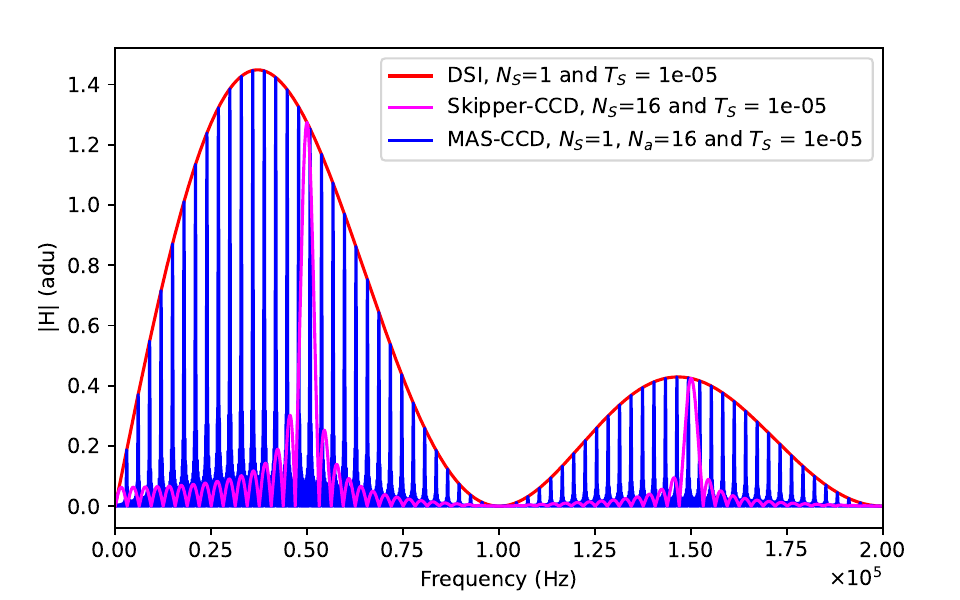}
    \caption{Amplitude comparison of the frequency response for DSI (red), the Skipper-CCD (magenta), and the MAS-CCD (blue) for the same number of total samples, i.e., $N_S$=1 for the MAS and $N_S$=16 for the Skipper. For the three $T_S=1e^-5$ is used.}
    \label{fig:rf_comparison1}
\end{figure}
Figure \ref{fig:rf_comparison1} illustrates the expected frequency response to the common noise of the MAS-CCD using $N_a=16$, $N_s=1$, and $T_s=1e-5$ compared to the expected one for the Skipper-CCD with a single sample ($N_a=N_s=1$) and with 16 samples in the output amplifier ($N_a=1$ and $N_s=16$). In the three cases the $T_s= 10\mu s$. The three operations filter out the common signals in different ways. The MAS-CCD plot in blue shows peaks with an amplitude given with an envelope given by the single samples single amplifier operation. Increasing the number of samples in a single amplifier (pink plot), the response still presents a single peak per lobe but it suppresses the lobes' amplitude at both sides frequencies. An interesting remark here is that noise lines in the common noise spectrum, for example, added by the readout system or poor grounding connections affecting all the amplifiers, could pass unfiltered when comparing its contribution from a single amplifier to a multiple amplifier solution.

\section{NRE recursive algorithm}
\label{appendix_GRE}
NRE recursive algorithm that was used to compare with the analytical calculations in Sec. \ref{sec:gre}. In this complete and fully functional example, the code functions are implemented for the case when the NRE is not a constant. The algorithm also provides a way to implement NRI values which are dependent on the charge in the sense node. The following code produces the measured charge by the amplifiers when a line of the active region is read and 10 overscan pixels, with $N_a=16$. It is assumed that the pixels in the active region have 62000e$^-$ of charge. The code accounts for 28 pre-scan pixels, before the first amplifier in the chain, and 15 pixels between amplifiers.

\newpage
\begin{python}[caption={NRI implementation in python.}]
import numpy as np
import matplotlib.pyplot as plt

def pixelsAfterNStages(pix, Na, NaPix):
    # OUTPUT
    # measPix is a matrix with the measured pixels by each stage in the 
    # row dimension and pixel index in the column direction.
    # The size(measPix): [Na, len(pix)].
    
    # INPUT
    # Na: number of amplifiers in the output stage of the MAS-CCD
    # pix: row vector with the charge information of the pixels in the 
    # serial register before the first amplifier in the output stage.

    Npix = len(pix)
    measPix = np.zeros((Na, Npix + (Na-1)*NaPix))#here we also add 
    # the pixels in the growing pre-scan for each amplifier
    inPix = pix.copy()

    for j in range(Na):   # run over all amplifiers
        leftInSN_j = 0  # variable accounting for the charge left 
        #behind in the SN. It is assume to start with zero charge 
        #when the row starts
        
        # run over the pixels in the serial register
        outPix = np.zeros(Npix)
        for n in range(Npix):  
            # measured pixel with index n by the amplifier j
            print(n)
            print(inPix[n])
            iColIndex = j*NaPix #the index accounts for the prescan of 
            #each amplifier
            measPix[j, iColIndex + n] = inPix[n]

            # Node Removal Inefficiency calculation (NRI). epsilon is 
            # the NRI value as used in the text of the article.
            # Epsilon is calculated from the epsilon function 
            #(epsilonFunc) which takes the total charge in the sense 
            # node of the amplifier of the argument.
            # This function is useful when epsilon varying with the 
            # charge value are needed.
            # If epsilon is the same for all pixel values, then this 
            # function epsilonFunc returns always the same value.
            epsilon = epsilonFunc(inPix[n] + leftInSN_j)
            movingForward = (inPix[n] + leftInSN_j) * (1 - epsilon)
            leftInSN_j = (inPix[n] + leftInSN_j) * epsilon
            # the pixel exits with charge:
            outPix[n] = movingForward
        inPix = outPix
    return measPix

def epsilonFunc(pix):
    # Example of a function to calculate the epsilon value for the 
    # case where this is dependent to the charge in the sense node.
    # The function takes the variable "pix" which is the amount of
    # charge in the sense node.
    # The implemented function is a linear extrapolation for epsilon 
    # between 3e-4 and 0.14 when the charge in the sense node ranges 
    # from 0 to 70,000 carriers.
    x = [0, 70e3]#[e-]
    y = [3e-4, 0.14]
    epsilon = np.interp(pix, x, y)
    return epsilon

#run the software from here
Npre = 28 #the number of pre-scan pixels before the first amplifiers
# in the chain
NpixAct = 256 #the width of the active area of the sensor
NpixOver = 10 #the desired number of pixels in the overscan
Q = 62000 #[e-] amount of charge collected by the pixels in the active
#region
Na = 16 #number of amplifiers in the serial register
NaPix = 15#number of pixels between output stages
pix = np.concatenate((np.zeros(Npre), np.ones(NpixAct)*Q, 
np.zeros(NpixOver)), axis=0) #vector of the pixels values in a row of
#the sensor considering prescan of the first amplifier and overscan
print(pix[30])
print(len(pix))
measPix = pixelsAfterNStages(pix, Na, NaPix) #matrix with the pixel
#information of measured by each amplifier

#plot the measured pixels by each amplifier
x = range(np.size(measPix,1))
for i in range(measPix.shape[0]):
    plt.plot(x, measPix[i], 'o--',label=f'Channel {i+1}',linewidth=1)
plt.xlabel('Column index in the output image')
plt.ylabel('Amount of charge (e-)')
plt.title('Pixel values measured by each amplifier')
plt.legend()
plt.show()
\end{python}
\end{spacing}
\end{document}